
\magnification=1200
\font\sc=cmcsc10
\font\Bsc=cmcsc10 scaled \magstep 1

\def\bx{{\bf x}}
\def\UN{{\bf 1}}
\def\HS{{\cal H}}

\def\xb{{\bf x}}

\def\mcg{mapping class group }
\def\tr{{\rm Tr}_q}

\def\TR{{\rm Tr}}
\def\c{\chi_p}
\def\o{\omega}

{\nopagenumbers
\null\vskip 3truecm

\centerline{\Bsc Mapping class group representations}\bigskip

 \centerline{\Bsc and generalized Verlinde formula}\bigskip\bigskip

\centerline{{\sc Peter B\'antay}\footnote{$^1$}{Partially supported
by OTKA grant F4010}}\bigskip

\centerline{Institute for Theoretical Physics\footnote{}{Email: {\tt
bantay@hal9000.elte.hu}}}\smallskip
\centerline{E\"otv\"os University, Budapest, Hungary}\bigskip

\centerline{and}\bigskip

\centerline{{\sc Peter Vecserny\'es}\footnote{$^2$}{Partially
supported by the Hungarian Research Fund, OTKA -- 1815}}\bigskip

\centerline{Fachbereich Physik -- Theoretische Physik\footnote{}{Email:
{\tt vecser@rhrk.uni-kl.de}}}\smallskip
\centerline{Universit\"at Kaiserslautern, Germany}\bigskip

\vskip 3truecm

\noindent Unitary representations of centrally
extended mapping class groups $\tilde M_{g,1}, g\geq 1$
are given in terms of a rational Hopf algebra $H$, and a related
generalization of the Verlinde formula is presented. Formulae expressing
the traces of \mcg elements in terms of the fusion rules, quantum dimensions
and statistics phases are proposed.

\vfill\eject}

\pageno=1

{\bf 1. Introduction}\medskip

Mapping class groups $M_{g,n}$ of $n$-holed genus $g$
oriented surfaces $\Sigma_{g,n}$ [1] occur as covariance
groups of (anti)holomorphic blocks of
$n$-point genus $g$ correlation functions in a RCFT [2].
 From the $M_{g,n}$-representations given in [2] on a peripheral
basis of vertex operators it is clear that only the superselection
structure [3] of the corresponding chiral RCFT is involved in
the game. However the superselection structure of a chiral
unitary RCFT can also be extracted from the DHR-endo\-morph\-isms [4]
of the underlying chiral observable algebra [5] and can be
encoded in the global symmetry algebra $H$: the unitary representations
of the global symmetry algebra and the DHR-endo\-morph\-isms
of the observable algebra form equivalent braided
monoidal rigid $C^*$-categories by definition [6] and the
common monodromy matrix, which is connected to a twofold braiding,
satisfies a non-degeneracy
condition [7, 8]. Therefore one should be able to derive the same
$M_{n,g}$ representations from the corresponding global symmetry
algebra $H$ itself.

In [8] rational Hopf algebras (RHA) were proposed as global
symmetry algebras of RCFT-s and the categorical equivalence
of RHA representations and DHR-endo\-morph\-isms of a rational QFT was
shown in [9]. In special cases of RHAs, namely, when $H$ is the
(possibly deformed) double ${\cal D}(G)$ of a finite group $G$,
the $M_{g,n}$ representations\footnote{$^1$}{For recent results in
the case of a general quantum double see [10], and for a more categorical
and knot theoretical settings see [11] and [12], respectively.}
were already given in [13] and meeting our expectations one can
extend this result to a general RHA.
Here we give a short summary of our results [14] on unitary
representation of the centrally extended mapping class groups
$\tilde M_{g,1}$ in terms of RHAs. It was inspired by the
geometrical presentation of $M_{g,1}$ on admissible tangles
given in [12].

As a byproduct of the genus one case $M_{1,1}$,
generalized fusion matrices $\{ N_p(t), t\in\hat H\}$
labelled by the irrep $t$ of the hole are found, that satisfy the usual
fusion algebra and are diagonalized by the $S(t)$-matrices, hence
can be expressed by a generalized Verlinde formula.

The sequence of mapping class group representations associated
in this way to the RHA should obey very specific consistency
conditions. We propose to investigate these conditions through
the study of the traces of the \mcg transformations, and give a
set of (conjectural) formulae expressing these traces in terms
of the usual data characterizing the RHA. We also derive a
curious result about the asymptotic (large genus)
behaviour of the traces.

\bigskip
{\bf 2. Unitary representations of  $\tilde M_{g,1}$}
\medskip

Let $U$ and $A$ denote the universal and the left
adjoint representations of the RHA $H$, i.e.
$$U=\bigoplus_{p\in\hat H} D_p,\qquad A=\bigoplus_{p\in\hat H}
D_p\times D_{\bar p},\eqno(1)$$
where $\hat H$ is the finite set of unitary equivalence classes of
irreducible $H$-representations, $D_p$ is a representative of the
class $p\in\hat H$, the bar indicates contragredient representations
and $\times$ denotes the product of representations. Let
$I(g,n)\equiv (D(g,n)\, \vert\, D(g,n))$ denote the
${}^*$-algebra of self-intertwiners of the product
representation $D(g,n):=A\times \dots\times A\times U\times
\dots\times U$ containing $g$  $A$ and  $n$ $U$ factors
respectively.\footnote{$^1$}{Although for a precise definition of
$D(g,n)$ we have to prescribe a bracketing in the
product since $H$ is not necessarily coassociative, all of the
bracketings lead to canonically equivalent representations, hence
canonically equivalent self-intertwiner spaces.}
$I(g,n)$ is a direct sum of full matrix algebras
$$I(g,n)=\bigoplus_{r\in\hat H} I_r(g,n)\equiv
  \bigoplus_{r\in\hat H} (D(g,n)\,\vert D_r)(D_r\,\vert\,
  D(g,n)),\eqno(2)$$
labelled by the intermediate irreducible represenation of $H$.
The elements of $M_{g,n}$ will be represented by unitary elements
from $I_0(g,n)$, where $0$ refers to the trivial representation.
$I_0(g,n)$ acts on the Hilbert space $\HS_0(g,n)\equiv
(D(g,n)\vert D_0)$ of intertwiners
by left multiplication. If $C_{pq}^{r\alpha}\in
(D_p\times D_q\vert D_r);\, p,q,r\in\hat H, \alpha=1,\ldots
,N_{pq}^r$ are the basic intertwiners, i.e.
$$C_{pq}^{r\alpha*}C_{pq}^{r'\alpha'}=\delta_{rr'}\delta_{\alpha
  \alpha'}D_r(\UN),\qquad \sum_{r\in \hat H}
  \sum_{\alpha=1}^{N_{pq}^r}C_{pq}^{r\alpha}C_{pq}^{r\alpha*}=
  (D_p\times D_q)(\UN),\eqno(3)$$
then an orthonormal basis of $\HS_0(g,n)$, the so-called path
basis, can be given by chains of basic intertwiners
$$\vert{\bf x}\rangle =C_{0p_1}^{c_1}C_{c_1\bar p_1}^{s_1\sigma_1}
  C_{s_1p_2}^{c_2\gamma_2}C_{c_2\bar p_2}^{s_2\sigma_2}\ldots
  C_{s_{g-1}p_g}^{c_g\gamma_g}C_{c_g\bar p_g}^{s_g\sigma_g}
  C_{s_gt_1}^{r_1\tau_1}C_{r_1t_2}^{r_2\tau_2}\ldots
  C_{r_{n-1}t_n}^0,\eqno(4)$$
labelled by the multiindex ${\bf x}=\{
p_i,c_i,s_i,\gamma_i,\sigma_i; t_j,r_j,\tau_j\vert i=1,\ldots ,g;
k=1,\ldots ,n\}$. A basis element can be visualized as the labelled
skeleton of a $\Sigma_{g,n}$ surface cutted along the
$d_i$ $(i=1.\ldots, g$) curves of Fig. 1. The irreps $p_i,\bar
p_i, t_k$  flow in the holes, the irreps $c_i,s_i,r_k$ label the
intermediate channels.

In every RHA there is a central unitary balance element
$b=S(R_2^*\varphi_1\lambda^*)l^*R_1^*\varphi_2lS(\rho\varphi_3)$
expressed by universal elements of $H$, which obeys the properties [9]
$$R_{21}R_{12}=(b^*\otimes b^*)\Delta(b),\qquad b=S(b),\eqno(5)$$
where $R\equiv R_{12}$ is the universal $R$-matrix of $H$ and $S$
denotes the antipode. The pure phases
$\omega_p=\omega_{\bar p}, p\in\hat H$ that appear in the central
decomposition of $b$ are called statistics phases.
The universal elements $r,l\in H$
lead to the statistics (or quantum) dimensions $d_p\geq 1$ of the
irreps :  $d_p=\TR\, D_p(rr^*)=\TR\, D_p(l^*l)$,
which satisfy the fusion rule algebra $d_pd_q=\sum_r N_{pq}^r d_r$ and
the equality $d_p=d_{\bar p}$. They can have non-integer values only in
case of unit-nonpreserving coproducts. In case of RHAs the quantity
$\sigma:=\sum_{p\in\hat H}\omega_p^*d_p^2$ satisfies
$\vert\sigma\vert^2=\sum_{p\in\hat H}d_p^2$ and defines the `central
charge' of the RHA: $c=(8/ 2\pi i)\log{\sigma^*/\vert\sigma\vert}\in [0,8)$.
Generalized $6j$-symbols, $F^{(pqr)_t}_{\alpha u\beta,\gamma v\delta}$,
of a RHA are given [15] in terms of
the basic intertwiners and the associator $\varphi$ of $H$ by
$$F^{(pqr)_t}_{\alpha u\beta,\gamma v\delta}\cdot
D_t(\UN)\delta_{tt'}=
  C_{ur}^{t'\beta *}C_{pq}^{u\alpha *}
  (D_p\otimes D_q\otimes D_r)(\varphi)
  C_{qr}^{v\gamma}C_{pv}^{t\delta}.\eqno(6)$$
They are unitary matrices in the lower multiindices describing
the change of the orthonormal product bases of basic intertwiners.
An appropriate choice of basic intertwiners [15] leads to a
simple form of them together with $S_4$ symmetry for the correctly
normalized symbols.

Let's turn to the \mcg representations.
It is known [1] that $M_{g,1}$ is generated by Dehn twists
$a_i, b_i, d_i, e_i, i=1,\ldots, g$ around the corresponding curves in Fig. 1.
and a presentation can be given in terms of
$a_i,b_i,\, i=1,\ldots ,g$ and $e_2$ [16, 12]. Knowing
the geometrical realization of $M_{g,1}$ by the isomorphic group, $T_{2g}$
of certain equivalence classes of admissible tangles [12], a central extension
$\tilde M_{g,1}:=\tilde T_{2g}$ of $M_{g,1}$ can be easily obtained by
splitting
up equivalence classes in $T_{2g}$: the central generator of $z$, which has
been the $K_1$-equivalence in [12],  will correspond to an insertion
(elimination)
of a $\pm 1$ ($\mp 1$) framed separated unknot into (from) a tangle diagram.
On the level of tangles it clearly
defines a central generator, which in terms of a RHA  is represented by
multiplication by the phase $\exp(\pm 2\pi ic/8)$.
Now $\tilde M_{g,1}$ has a presentation in terms of
the generators $a_i,b_i,\, i=1,\ldots ,g;\, e_2, z$ together with
the relations
$$b_ia_ib_i=a_ib_ia_i,\qquad b_ia_{i+1}b_i=a_{i+1}b_ia_{i+1},
  \qquad b_2e_2b_2=e_2b_2e_2,\eqno(7a)$$
and every other pair of generators commute,
$$\eqalign{(a_2b_1a_1)^4=z^4ke_2k^{-1}e_2,\qquad
  k&=b_2a_2b_1a_1^2b_1a_2b_2,\cr
  (a_3a_2a_1)^{-1}g_1g_2e_2=we_2w^{-1},\qquad
  g_2&=(b_2a_3a_2b_2)e_2(b_2a_3a_2b_2)^{-1},\cr
  g_1&=(b_1a_2a_1b_1)g_2(b_1a_2a_1b_1)^{-1},\cr
  w&=b_3a_3b_2a_2b_1a_3^{-1}b_3^{-1}g_2
    b_3a_3a_1^{-1}b_1^{-1}a_2^{-1}b_1^{-1}.\cr}  \eqno(7b)$$
Having constructed a homomorphism [14] --- suggested by the tangle
diagrams --- from $\tilde T_{2g}$ into unitary $I_0(g,1)$ intertwiners
of a RHA we obtain unitary representations of $\tilde M_{g,1}$.
Using the multiindex notation of (4) the
explicit form of this representation is as follows:

$$ Z_{\bx\bx'}=\delta_{\bx\bx'}\exp(2\pi ic/8),\qquad
    D(i)_{\bx\bx'}=\delta_{\bx\bx'}\omega_{p_i},\qquad
    E(i)_{\bx\bx'}=\delta_{\bx\bx'}\omega_{c_i},$$
$$ A(i)_{\bx\bx'}=\delta_{\bx\bx'}
  (\sigma_{i-1}s_{i-1}\gamma_{i-1})
  \sum_{q,\alpha,\beta}
  F^{(c_{i-1}\bar  p_{i-1}p_i)_{c_i}}_{\sigma_{i-1}
  s_{i-1} \gamma_{i-1},\alpha q\beta}\cdot\omega_q\cdot
  F^{(c_{i-1}\bar p_{i-1}p_i)_{c_i}*}_{\alpha q\beta,
  \sigma_{i-1}^\prime s_{i-1}^\prime\gamma_{i-1}^\prime},
  \eqno(8)$$
$$ B(i)_{\bx\bx'}=\delta_{\bx\bx'}(p_i\gamma_ic_i\sigma_i)
   \exp(2\pi ic/8)\omega_{p_i}^*\omega_{p_i^\prime}^*
  \sum_{q,\beta\atop\alpha,\alpha'}
  F^{(s_{i-1}p_i\bar p_i )_{s_i}}_{\gamma_i c_i\sigma_i,\alpha q\beta}
  {S(\bar q)^*}_{p_i\alpha}^{p_i^\prime\alpha'}
  F^{(s_{i-1}p_i^\prime \bar p_i^\prime )_{s_i}*}_{\alpha' q\beta,
  \gamma_i^\prime c_i^\prime\sigma_i^\prime},$$
where $\delta_{\bx\bx'}(\,\cdot\, )$ means the Kronecker delta in
the multiindex except the subindices in its argument.

We note that the given
representation does not involve intertwiners that change the
representation $t$ of the hole therefore the representation space
decomposes as $\HS_0(g,1)=\oplus_t\HS_0(g;t)$.

As an illustration of the $\tilde M(g,1)$ representations (8)
let us consider the case of the Lee-Yang fusion rules:
$$N_0=\left(\matrix{1&0\cr 0&1\cr}\right)\qquad
  N_1=\left(\matrix{0&1\cr 1&1\cr}\right).$$
There are two inequivalent RHAs having these fusion
rules [15], they correspond to the unitary level one $G_2$ and $F_4$
KM-algebra representations: the common statistical dimensions are $d_0=1,
d_1\equiv d=(1+\sqrt{5})/2$ and the statistics phases are
$\omega_0=1,\omega_1=\omega^4$, with $\omega=\exp(\pm 2\pi i/10)$.
The corresponding central charges are $14/5$ and $26/5$.

For a simple form of the nondiagonal $\tilde M_{g,1}$ generators
in (8) we have decomposed the unit into orthogonal projections:
$P^{c_{i-1}\bar p_{i-1}p_ic_i}_{[s_{i-1}]}$
for the generators $A(i)$ and $P^{s_{i-1}s_i}_{[p_ic_i]}$
for the generators $B(i)$. The upper indices (with fixed
admissible irrep values) show the projections to a subspace within
the generators can have nondiagonal matrix elements according to the
corresponding argument for the Kronecker-$\delta$ in (8). The lower indices
show the rank of these projections by listing the admissible irrep
values, i.e. the basis for this subspace. The
order of the list shows how the connected matrix is understood.
$$Z_{\xb\xb'}=\delta_{\xb\xb'}\exp 2\pi i {\pm 7\over 20}\qquad
  D(i)_{\xb\xb'}=\delta_{\xb\xb'}\omega_{p_i}\qquad
  E(i)_{\xb\xb'}=\delta_{\xb\xb'}\omega_{c_i}$$
$$\eqalign{A(i)_{\xb\xb'}&=\delta_{\xb\xb'}(s_{i-1})\Bigl\{
   P^{0000}_{[0]}+P^{0110}_{[1]}+P^{1001}_{[1]}+
  {1\over d^2}\left(\matrix{1+\omega d&\sqrt{d}(1-\omega)\cr
                       \sqrt{d}(1-\omega)&\omega+ d\cr}\right)
    P^{1111}_{[0,1]}\cr
  &+  \omega\left( P^{0011}_{[0]}+P^{0101}_{[1]}+P^{0111}_{[1]}+
     P^{1010}_{[1]}+P^{1011}_{[1]}+P^{1100}_{[0]}+P^{1101}_{[1]}+P^{1110}_{[1]}
    \right)\Bigr\}\cr}$$
$$\eqalign{B(i)_{\xb\xb'}&=\delta_{\xb\xb'}(p_i,c_i)\Bigl\{
   P^{01}_{[11]}+P^{10}_{[11]}
  +{\exp 2\pi ic/8\over\vert\sigma\vert}\left(\matrix{1&\omega^{-1}d\cr
   \omega^{-1}d&\omega^3\cr}\right)P^{00}_{[00,11]}\cr
    &+{\omega^{-1}\exp 2\pi ic/8\over\vert\sigma \vert }
  \left(\matrix{\omega&1&\sqrt{d}\cr
      1&d^{-2}(\omega^4\mp id\vert\sigma\vert)&
              d^{-3/2}(\omega^4\pm i\vert\sigma\vert) \cr
            \sqrt{d}&d^{-3/2}(\omega^4\pm i\vert\sigma\vert)
      &d^{-2}(\omega^4d\mp i\vert\sigma\vert)\cr}\right)
     P^{11}_{[01,10,11]}\Bigr\}\cr}$$
In case of mapping class group generators $A(1),B(1)$
irreps with index $i=0$ are always understood as the trivial ones:
$p_0=s_0=c_0=0$. Moreover, $c_1=p_1$ and $s_g=\bar t$ are always valid.
An explicit check of (7) can be easily performed by using
that $d=\omega+\omega^{-1}$ and $d^2-d-1=0$.

\bigskip

{\bf 3. Representation of $M_{1,1}$ and the generalized Verlinde formula}
\medskip

An equivalent  (redundant) presentation of $M_{1,1}$ to (7) can be
given by
$$\langle T,S,R\,\vert (ST)^3=S^2, S^4=R^{-1}\rangle,\eqno(9)$$
where $R$ has the geometrical meaning of a twist around
the hole of the surface $\Sigma_{1,1}$.
The correspondence between the $M_{1,1}$ generators in (7) and
(9) is as follows
$$S=(ABA)^{-1}\exp(2\pi ic/8),\qquad T=A\exp(-2\pi ic/24).
  \eqno(10)$$
As we have already mentioned the representation space decomposes
according to the irrep $t$ flowing through the hole,
$\HS_0(1,1)=\oplus_t\HS_0(1;t)$, and
the \mcg transformations decompose into a direct sum accordingly.
In these subspaces the path basis is characterized by the pair
$p,\alpha$ with $\alpha=1,\dots, N_{p\bar p}^{\bar t}$ and the
generators are represented by the unitary matrices
$$T(t)_{p\alpha}^{\, p^\prime\alpha^\prime}=\delta_{pp^\prime}
  \delta_{\alpha\alpha^\prime}\omega_p  e^{-2\pi i{c\over
24}},\qquad
  S(t)_{p\alpha}^{\, p^\prime\alpha^\prime}={1\over\vert\sigma\vert}
  \sum_r{\omega_r\over\omega_p\omega_{p^\prime}}
  d_rN_r(t)_{p\alpha}^{\, p^\prime\alpha^\prime},\eqno(11)$$
where the generalized fusion matrices $N_r(t)$ read as
$$N_r(t)_{p\alpha}^{\, p^\prime\alpha^\prime}=
  \sum_{\beta=1}^{N_{rp}^{p'}}
  F^{(p'\bar r\bar p)_{\bar t}}_{\beta p\alpha,\beta \bar p'\alpha'}.
  \eqno(12)$$
They satisfy the usual fusion algebra
$$N_p(t)N_q(t)=\sum_r N_{pq}^r N_r(t) \qquad t\in\hat H,\eqno(13)$$
and are diagonalized by the corresponding $S(t)$ transformations:
$$N_q(t){\bf s}^{p\alpha}(t)={S(0)_q^p\over S(0)_0^p}\cdot
{\bf s}^{p\alpha}(t),\eqno(14)$$
where ${\bf s}^{p\alpha}(t)$ are the column vectors of $S(t)$.
Therefore a generalized Verlinde formula holds:
$$N_r(t)_{p\alpha}^{\, p^\prime\alpha^\prime}=
  \sum_q\sum_{\beta=1}^{N_{q\bar q}^t}
  {S(0)_r^qS(t)_{p\alpha}^{\, q\beta}
  {S(t)^*}_{\, q\beta}^{p'\alpha'} \over S(0)_0^q}.\eqno(15)$$
Due to the $S_4$-symmetry of the
normalized $6j$-symbols the generalized fusion coefficients
obey also the properties
$$  N_r(t)=N_{\bar r}(t)^*=\overline{N_r(\bar t)}=
  N_{\bar r}(\bar t)^T,\eqno(16)$$
where star, overline and ${}^T$
indicate adjoint, complex conjugate and transposed matrices.

As an illustration of the generalized fusion matrices and the
Verlinde formula let us consider the $(7,2)$ fusion rules:
$$N_0=\left(\matrix{1&0&0\cr 0&1&0\cr
0&0&1\cr}\right),\quad
  N_1=\left(\matrix{0&1&0\cr 1&0&1\cr 0&1&1\cr}\right),\quad
  N_2=\left(\matrix{0&0&1\cr 0&1&1\cr 1&1&1\cr}\right).$$
There are two inequivalent RHAs corresponding these fusion rules [15]:
the common statistical dimensions are $d_0=1, d_1=d, d_2=d^2-1$ with
$d=2\cos(\pi/7)$ and the statistics phases are $\omega_0=1,\omega_1=
\omega^2, \omega_2=\omega^{10}$ with $\omega=\exp(\pm i\pi/7)$. The
corresponding central charges are $48/7$ and $8/7$ and
$\vert\sigma\vert=d^2+d-2$. The generalized fusion matrices read as
$$N_0(1)=1,\qquad N_1(1)={1\over d^2-1},\qquad  N_2(1)=-{d\over d^2-1}.$$
$$N_0(2)=\left(\matrix{1&0\cr  0&1\cr}\right),\quad
  N_1(2)=\left(\matrix{0&-{1\over\sqrt{d}}\cr
            -{1\over\sqrt{d}}&-{d\over d^2-1}\cr}\right),\quad
  N_2(2)=\left(\matrix{-{1\over d^2-1}&{\sqrt{d}\over d^2-1}\cr
            {\sqrt{d}\over d^2-1}&{1\over d^2+d}\cr}\right).$$
The corresponding $S(t)$-matrices are
$$S(0)={1\over \vert\sigma\vert}
  \left(\matrix{1&d&d_2\cr d&-d_2&1\cr d_2&1&-d\cr}\right), \quad
   S(1)=\omega^{10},\quad
   S(2)={\omega^{-3}\over\vert\sigma\vert}
     \left(\matrix{\omega^{-1}+\omega^2 &\sqrt{d}(\omega+1)\cr
            \sqrt{d}(\omega+1)&-\omega^{-1}-\omega^2\cr}\right).$$
That the $S(t)$-matrices diagonalize the corresponding fusion rules and
$S(t)^4=\omega_t^{-1}N_0(t)$ can be easily checked by noting that
$d=\omega+\omega^*$ and $d^3-d^2-2d+1=0$.

We close this Section by mentioning that if we had allowed the
presence of a nontrivial degenerate sector $p\in\hat H,p\not=0$
--- when, by definition, all monodromies of $p$ are trivial,
i.e. $(D_p\otimes D_q)(R_{21}R_{12}) =(D_p\times D_q)(\UN)$ for
all $q$ --- then the $S(0)$ matrices would have been degenerate
and the $S(t),\ t\not=0$ matrices would have even contained identically
zero rows corresponding to the degenerate sector.
This means that a nontrivial sector should be
braided at least with one of the others in order to obtain a
modular structure. Sectors having completely permutation
statistics are excluded.

\bigskip{\bf 4. Trace formulae}\medskip

As we have seen in the previous sections, the knowledge of the RHA associated
to a given theory enables us to compute explicitly the relevant \mcg
representations. But for some applications, e.g. the classification of
RCFT-s or the study of the higher genus aspects of string theory, this is not
really what is needed. In these problems the interesting question is to find
out what kind of relations exist between the different higher genus
characteristics
of a given theory, for example between the (linear equivalence classes of)
\mcg representations. That there should be some relations is clear from the
fact that all these representations are determined by a finite set of data,
the $6j$-symbols and statistics phases of the RHA. Our task is to
make them explicit.

A natural strategy is to try to express the above mentioned relations in terms
of the traces of the \mcg transformations. We shall illustrate this idea first
in the case of the one-holed torus, and we shall comment on the general case
later.

A word of caution is in order here. All of the results to be presented in this
Section have been derived in the context of orbifold models, i.e. the
corresponding RHAs are (possibly deformed) doubles of a finite group, where the
availability of group theory techniques made possible the computations. We
have not been able to prove them for general RHAs, so the
status of the following results is only conjectural. Nevertheless, we have
checked
them numerically in a variety of models, including those with $Z_2$, $Z_3$,
Ising, Lee-Yang, and $(7,2)$ fusion rules, and in our opinion this strongly
supports their validity in general.

Let's turn to the results.  It is more convenient to work with
the class functions $\chi_p$ instead of the partial traces $\TR_t$ on
$\HS_0(1;t)$.
The former are defined for $X\in M_{1,1}$ via the (invertible) rule
$$\chi_p(X)=S_{0p}\sum_t\bar S_{pt}\TR_t(X),\eqno(20)$$
where we use $S_{pt}=S(0)_p^t$ for short.
A simple observation is that the $\c$-s are normalized, i.e.
$\c({\bf 1})=1$. Moreover, $\c(X^{-1})=\bar\chi_{\bar p}(X)$.

The first interesting result is that for many elements $X\in M_{1,1}$,
the quantity $\chi_p(X)$ may be expressed in terms of the fusion rules,
the quantum dimensions and the statistics phases of the sectors.
Here is a sample (with the notation  $\eta=\exp(-i\pi{c\over 12})$):

\bigskip

\vbox{\offinterlineskip\settabs 5\columns
\+&{\vrule height+10pt}$\displaystyle\qquad\qquad\qquad\qquad\chi_p$\cr
\+&{\vrule height+6pt}\cr\hrule
\+$T^n$&{\vrule height+22pt}$\displaystyle\qquad\qquad\eta^n\sum_q\vert
S_{pq}\vert^2\o_q^n$\cr
\+$S$&{\vrule
height+22pt}$\displaystyle\qquad\qquad{\eta^6\over\vert\sigma\vert^3}
\sum_{q,r}N_{pq}^rd_pd_qd_r{\o_p^4\o_r^4\over\o_q^2}$\cr
\+$S^2T^n$&{\vrule
height+22pt}$\displaystyle\qquad\qquad{\eta^n\over\vert\sigma\vert^4}
\sum_{q,r,s,t}N_{pq}^tN_{rs}^td_pd_qd_rd_s
{\o_p^2\o_r^2\over\o_q^2\o_s^2}\o_t^n\hskip 2.7cm (21)$\cr
\+$ST$&{\vrule
height+22pt}$\displaystyle\qquad\qquad{\eta^7\over\vert\sigma\vert^3}
\sum_{q,r}N_{pq}^rd_pd_qd_r{\o_p^6\o_r^3\over\o_q^2}$\cr
\+$\left(ST\right)^2$&{\vrule height+22pt}$\displaystyle\qquad\qquad
{\eta^8\over\vert\sigma\vert^4}\sum_{q,r,s,t}N_{pq}^tN_{tr}^sd_pd_qd_rd_s
{\o_p^3\o_q^3\o_r^3\over\o_s}$\cr}
\bigskip
\medskip

Let's observe that the above
formulae lead to  non-trivial relations even for the partial trace $\TR_0$,
that is for the closed torus, e.g. for the transformation $S$ we get
$$\sum_p\chi_p(S)\equiv{\rm Tr}_0(S)={\eta^6\over\vert\sigma\vert^3}
\sum_{p,q,r}N_{pq}^rd_pd_qd_r{\o_p^4\o_r^4\over\o_q^2},\eqno(22)$$
the last equality being far from trivial.

The fact that the transformation $S^4$ is nothing but a Dehn-twist around the
hole implies
$$\chi_p(S^4X)=\eta^3\sum_q{S_{pq}d_p\o_p\o_q\over d_q}\chi_q(X).
   \eqno(23)$$

Two more important properties of the class functions $\chi_p$ will be of
use later. The first is that if $d_r=1$ and $N_{pr}^q=1$, then
$\chi_p=\chi_q$, in particular $\chi_p=\chi_0$ for all irreps
$p$ whose quantum dimension is 1.
The second one is the following interesting factorization property:
$$\chi_0\left(T^kS^{-1}T^nS\right)=\chi_0\left(T^k\right)\chi_0\left(
T^n\right).\eqno(24)$$

Up to now we have been concerned with the mapping class group $M_{1,1}$ of
the one-holed torus, but clearly the above results may be applied to
investigate the higher genus mapping class groups.
This is so because sewing allows one to imbed naturally (in many different
ways) $M_{1,1}$ in $M_{g,0}$, and any such imbedding allows us to view a
transformation $X\in M_{1,1}$ as an element $\hat X$ of $M_{g,0}$.
A simple argument shows that
$${\rm Tr}(\hat X)=\sum_{p,q}S_{0p}^{1-2g}
S_{qp}\tr(X)=\sum_pS_{0p}^{2-2g}\chi_p(X),\eqno(25)$$
where the trace on the lhs. is over the space of genus $g$ characters.
Note that $(25)$ implies that asymptoticaly
$${\TR(\hat X)\over\TR({\bf 1})}\to \chi_0(X)\qquad{\rm as}\qquad g\to\infty
,\eqno(26)$$
which together with $(24)$ leads to the following curious result :

{\it If $\delta_1,\delta_2 \in M_{g,0}$ are Dehn-twists around simple closed
curves whose linking number is one, then in the limit $g\to\infty$
the trace of their product factorizes up to a constant of proportionality
equal to the trace of the identity transformation.}

\bigskip{\bf 5. Concluding remarks}\medskip

The morale of our work is that the modular geometry of a CFT is just a
reflection of its superselection structure. One of the advantages of the
algebraic approach developed in this paper is that it makes easier the
deeper study of this connection. We feel that the trace formulae presented
in the last section could be a first step in this direction. Of course,
much more has to be done to achieve this goal.

We hope that our work would be useful in the study of field theories
defined on a space-time of nontrivial topology as well as in a better
understanding of the nonperturbative aspects of string theory.

As a final remark we note that although the emergence of \mcg transformations
seems to be natural in the
context of conformal field theories, one should be able to give a meaning of
these transformations when a RHA arises as a global symmetry of a QFT without
conformal invariance (e.g. certain lattice field theories [17]). We think that
Schroer's vacuum polarization picture [18] is a natural answer: \mcg
transformations describe the possible unitary selfintertwiners in the presence
of
spectator charges and vacuum splittings that `sum up' to the trivial sector. In
our
treatment the $M_{g,n}$ generators act exactly on this `spectator' space, on
$\HS_0(g,n)$, where the factors of universal and the left adjoint
representations
correspond to the spectator charges and the vacuum splittings, respectively.
\medskip

{\it Acknowledgements.} We would like to thank R. Rim\'anyi for the
numerous valuable discussions.

\bigskip{\bf References}
\medskip

\item{[1]} J.S. Birman, Contemp. Math. 78 (1989) 13 \smallskip

\item{[2]} G. Moore and N. Seiberg, Commun. Math. Phys.
123 (1989) 177\smallskip

\item{[3]} R. Haag, {\it Local quantum physics}, Springer 1992
\smallskip

\item{[4]} S. Doplicher, R. Haag and J.E. Roberts, Commun. Math.
Phys. 13 (1969) 1; 15 (1969) 173;
23 (1971) 199; 35 (1974) 49 \smallskip

\item{[5]} K. Fredenhagen, K.-H. Rehren and B. Schroer,
Rev. Math. Phys., special issue 113, (1992)\smallskip

 \item{[6]} S. Doplicher and J.E. Roberts, Ann. Math. 130
(1989) 75; Invent. Math. 98 (1989) 157; Commun. Math. Phys. 131
(1990) 51, \smallskip

\item{[7]} K.-H. Rehren, {\it Braid group statistics and their
superselection rules}, in Algebraic theory of superselection
sectors, ed. D. Kastler, World Scientific, p. 379, 1990
\smallskip

\item{[8]} P. Vecserny\'es, Nucl. Phys. B415 (1994) 557
\smallskip

\item{[9]} K. Szlach\'anyi and P. Vecserny\'es, {\it Rational Hopf
algebras as global symmetries of rational field theories} (to be
published)\smallskip

\item{[10]} T. Kerler,  Commun. Math. Phys. 168 (1995) 353, \smallskip

\item{[11]} V. Lyubashenko, J. Pure and Appl. Alg. 98 (1995) 279
\smallskip

\item{[12]} S. Matveev and M. Polyak, Commun. Math. Phys.
160 (1994) 537\smallskip

\item{[13]} P. B\'antay, Int. J. Mod. Phys. A9 (1994) 1443 \smallskip

\item{[14]} P. B\'antay and P. Vecserny\'es, (to be published)
\smallskip

\item{[15]} J. Fuchs, A. Ganchev and P. Vecserny\'es,
NIKHEF-H/94--05 preprint (to appear in
Int. J. Mod. Phys. A)\smallskip

\item{[16]} B. Wajnryb,  Israel. J. Math. 45 (1983) 157\smallskip

\item{[17]} K. Szlach\'anyi and P. Vecserny\'es,  Commun. Math. Phys.
156 (1993) 127\smallskip

\item{[18]} B. Schroer, preprint, hep-th 9405105 (1994)\smallskip
\vfill\eject

\input epsf
\epsfxsize=450pt
\epsfysize=200pt
\epsfbox{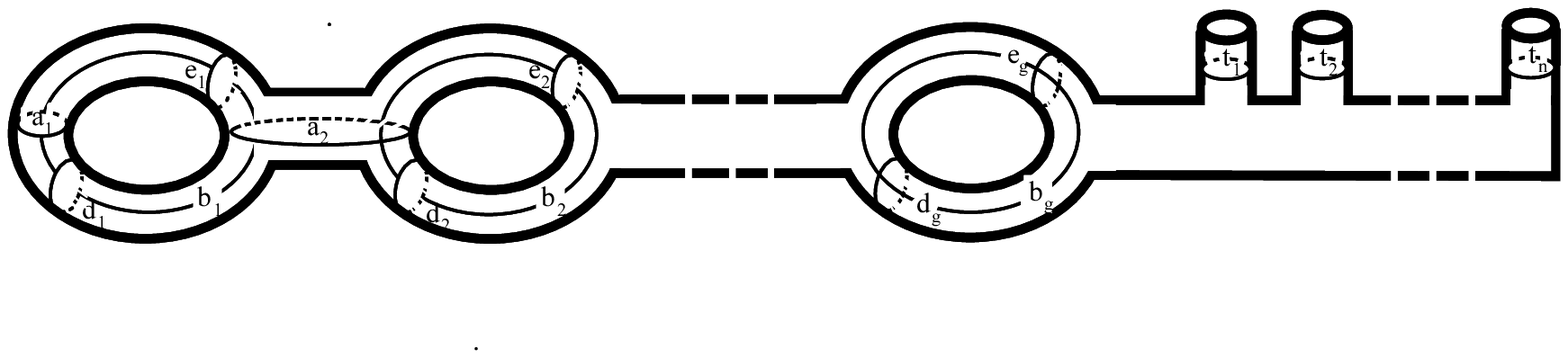}
\bye